\documentclass[a4paper, twoside, headsepline]{scrartcl}
\usepackage{graphicx}
\setcapindent{0em}
\graphicspath{{./} {./fig/}}

\begin{document}
\title{What's wrong with Phong \\
- Designers' appraisal of shading in CAD-systems}

\author{J\"org M. Hahn\thanks{Daimler AG, 
Design - Data and Models, 
Sindelfingen, Germany, 
joerg.hahn(at)daimler.com.
This paper was written 2006-2007.
The author thanks the designers at Mercedes-Benz
who were critical of their digital renderings,
and who helped with observations and comments.
The author gratefully acknowledges feedback and encouragement
from Konrad Polthier of Freie Universit\"at Berlin,
and from Roland Fleming of the Max Planck Institute for Biological Cybernetics
in T\"ubingen.
}}

\date{}

\maketitle

\begin{abstract} 
The Phong illumination model is still widely used in realtime 3D visualization systems.
The aim of this article is to document problems with the Phong illumination 
model that are encountered by an important professional user group, 
namely digital designers.  
This leads to a visual evaluation of Phong illumination,
which at least in this condensed form seems 
still to be missing in the literature.
It is hoped that by explicating these flaws, awareness about the limitations and
interdependencies of
the model will increase, both among fellow users, and among researchers and
developers.  
           
\end{abstract}

\section{Introduction}

The Phong illumination model,
due to its simplicity and ability to model a range of material appearances, 
has become one of the most widely used shader models in computer graphics.
Nevertheless, it is well known that it has certain shortfalls.

In the original paper \cite{Phong}  no physical justification was given
nor indeed intended.
Later it was noted that it is not physically plausible,
e.g Helmholtz reciprocity and energy conservation are not met \cite{Lewis}.
And in the perception literature, the physically unrealistic appearance has been noted,
e.g. 
\cite{ParkerChristouCunnings},
\cite{JohnstonCurran}, 
\cite{KoenderinkShading},
although it is still com\-monly used in psychophysical experiments.

And it is still predominant in CAD-systems.  The common
graphics interfaces, OpenGL and DirectX, employ the Phong illumination model.
And thus it is encountered
by most if not all users who create digital 3D models.

Industrial Design today makes heavy use of digital means to materialize ideas.
The final product is a real (physical) object.
In this respect,
design differs from other prominent areas of digital rendering,
e.g. the games and movie industries, 
where only the visual appearance of the digital model counts.

In particular in the automotive industries there is a well established
division of labour between designers\footnote{ 
The term \emph{design} refers to conceptual design or styling 
as opposed to engineering design.
}
and modellers.
Roughly speaking, the designer sketches ideas (mainly in 2D) and 
the modeller builds a 3D model,
either physical or digital.  
Then it is the task of the designer to look at, perceive and understand 
the model built by the modeller and then to 
refine his sketches to direct 
the further evolution of the model.

The appraisal of digital models and physical models is a vital skill
for a designer.
There is hardly any other user group that looks at digital models 
as carefully and critically as designers do\footnote{
It seems that the modellers have a much better understanding of their models.
But their understanding of the model is less based on the visual appearance. 
They rely much more on wire-frame display 
(iso-parameter lines) or diagnostics (e.g. section lines, isophotes).
And they interact longer and more intensively with the digital model.  
}.
Their observations and impressions can give valuable input to computer graphics
research. 

There seems to be surprisingly little research on how designers appraise digital
models.
An exception is \cite{Ferwerda},
which describes an experiment with color frogs, 
which are generic car-like shapes.
They found that the 
rendering method used had a significant effect on the ability to
discriminate color frogs that differed subtly in shape,
and global illumination rendering improved sensitivity to shape differences.

On the other hand, Greg Ward, the author of the 
Radiance lighting simulation and rendering system \cite{WardRadiance},
emphasizes the importance of the local illumination model
for realistic appearance of digital renderings
\cite{WardLocalIllum}, \cite{WardReal},
but suspects that this still seems to be poorly understood.

Occasional complaints of CAD modellers about the apparent form of shaded images
of their digital models,
and some surprise of designers when looking at a physical model milled from
digital data,
gave reason for an investigation of digital models and their display 
in the CAD-systems concerned.
In fact, a stark statement from an embarrassed designer,
\emph{the digital model looks like a cardboard box},
imposed the question,
what makes a nicely curved surface look like a cardboard box,
and thus gave the ignition. 

The CAD-systems employed were
Alias StudioTools V12 with its products AutoStudio and SurfaceStudio, 
CATIA V5R13, and ICEM Surf V4.3,
which are widely used in the automotive industries for modelling,
and an in-house system, DBView/veo, 
a real time visualization system based on OpenInventor
which is mainly used for design evaluation by designers.
In working sessions a designer often joins a modeller working in ICEM or Alias
and sees the display of these systems.
CATIA was added mainly due to its importance for automotive engineers.

It soon turned out that illumination was a main issue of concern,
and the question was,
how to evaluate or understand this concern.
So an idea emerged to conduct a hands-on experiment with some designers.
In the course of this investigations other flaws of the illumination
were found or re-discovered.
This led to a visual evaluation of Phong illumination,
which at least in this condensed form seems 
- more than 30 years after its origin -
still to be missing in the literature.

The aim of this article is to document the problems with the Phong model 
that are encountered by an important professional user group, 
namely digital designers.
It is hoped that by explicating these flaws, 
awareness about the limitations and interdependencies of
the model will increase, both among fellow users, and among researchers and
developers.  

This paper is organized as follows.
Section 2 shows an experiment with designers on shape perception with
different lighting conditions.
Section 3 looks on some (more or less known) 
effects of Phong illumination and highlights
that are visible to designers,
and discusses their impact on form appraisal.
Section 4 deals with the behaviour of light and material and related effects 
that are visible to and experienced by designers.
Finally, section 5 summarizes the conclusions.

\section{Form appraisal - an experiment with designers}
In a workshop with a group of automotive designers,
the designers were given some tasks involving digital images.
The images were presented on a powerwall.
The designers were all familar with digital presentations of digital models.

\subsection{Model selection - ball} 
\label{sec:ball}

\begin{figure}[htbp]
  \begin{minipage}[b]{.48\linewidth}
    \centering{
    (a) \includegraphics[width=.6\linewidth]{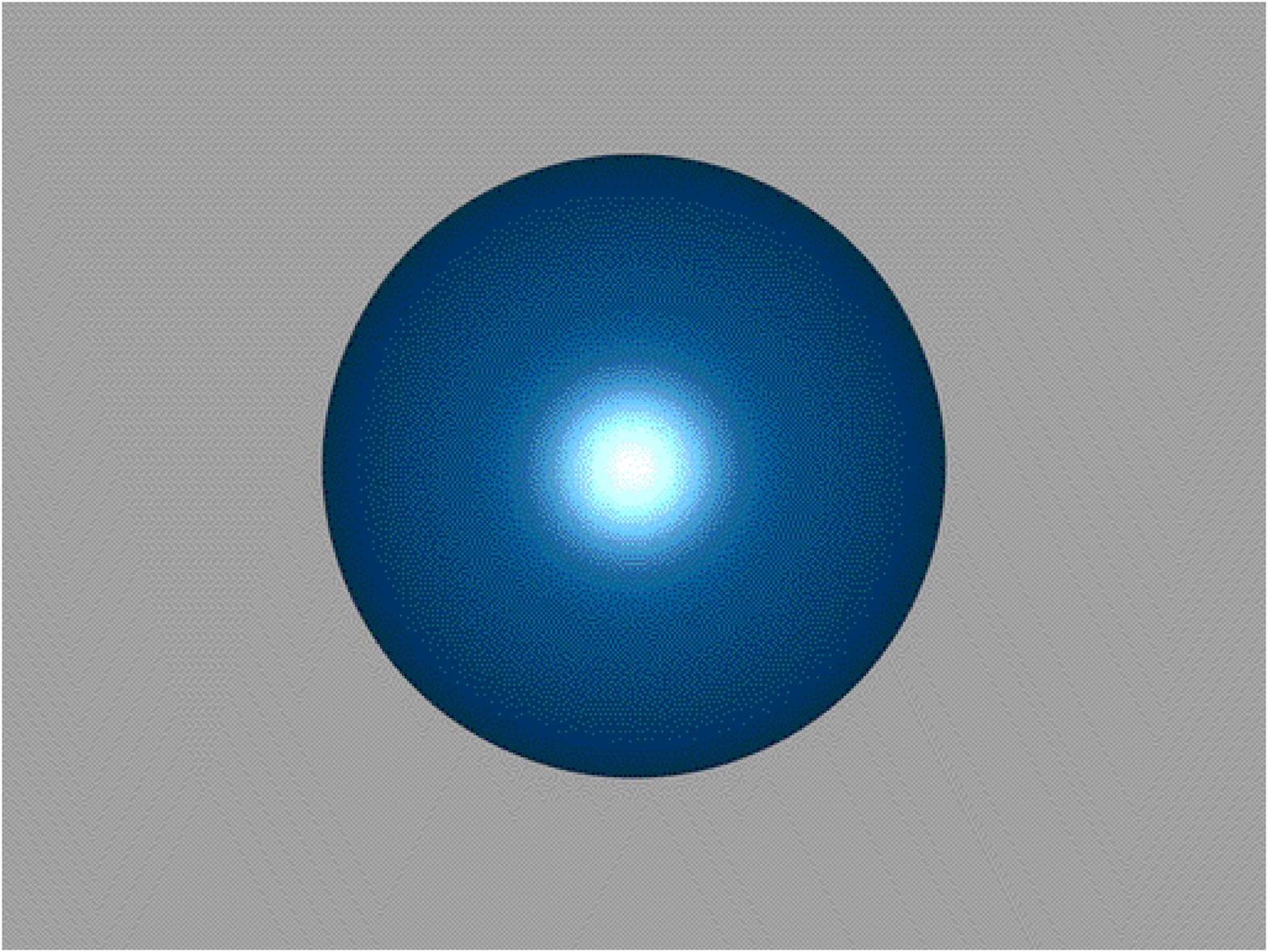} \\
    (b) \includegraphics[width=.6\linewidth]{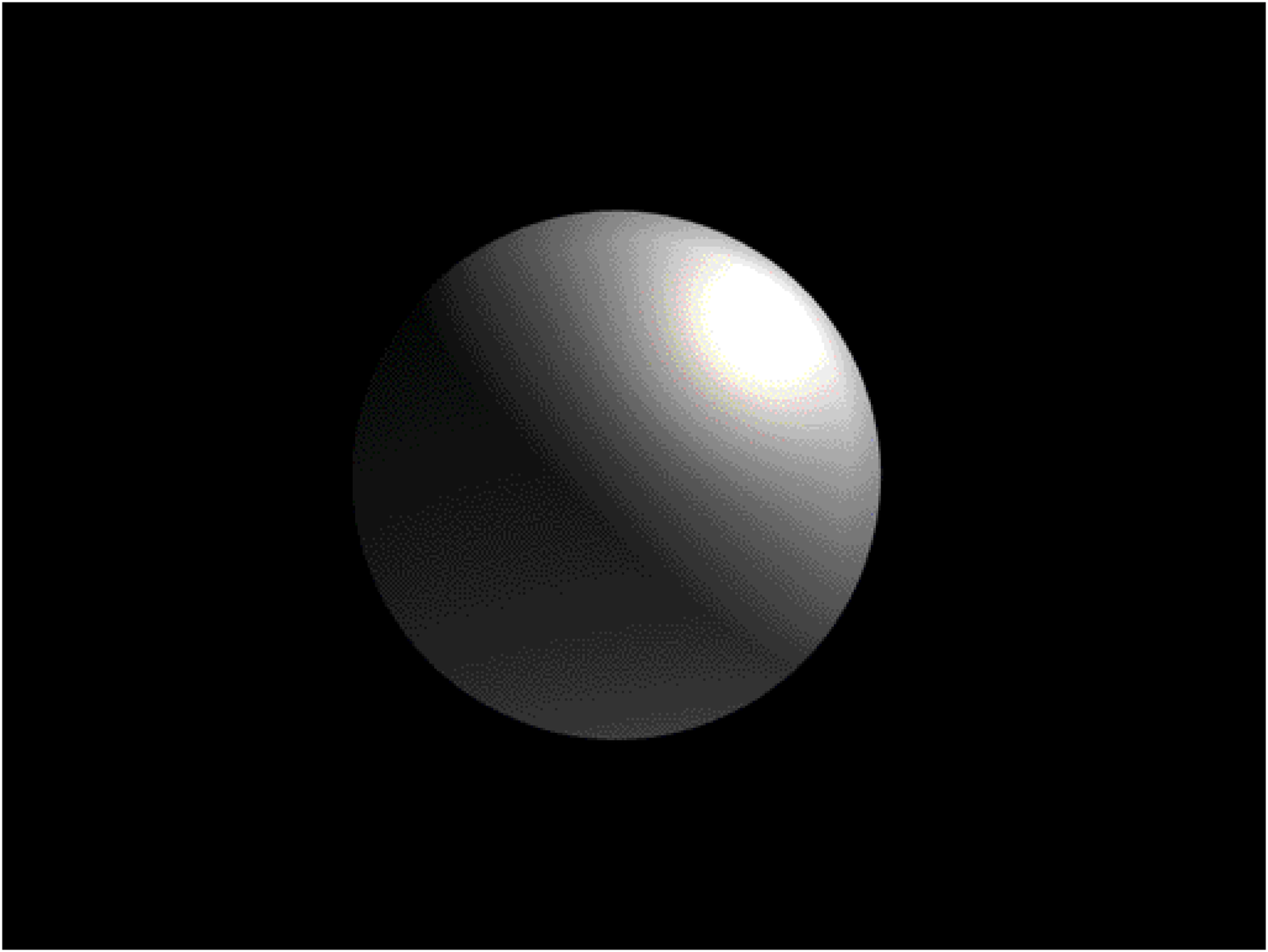} \\
    (c) \includegraphics[width=.6\linewidth]{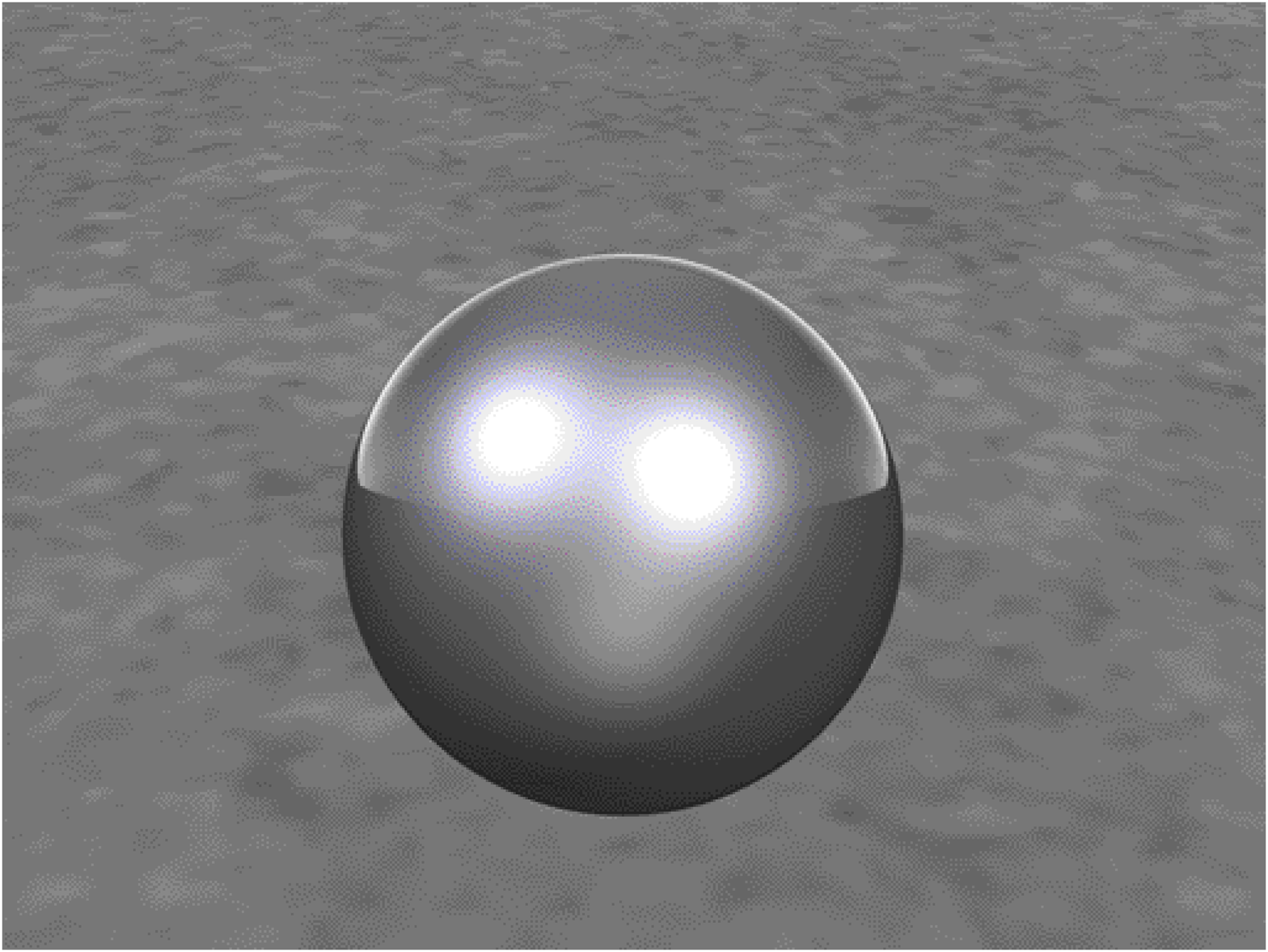} \\
    (d) \includegraphics[width=.6\linewidth]{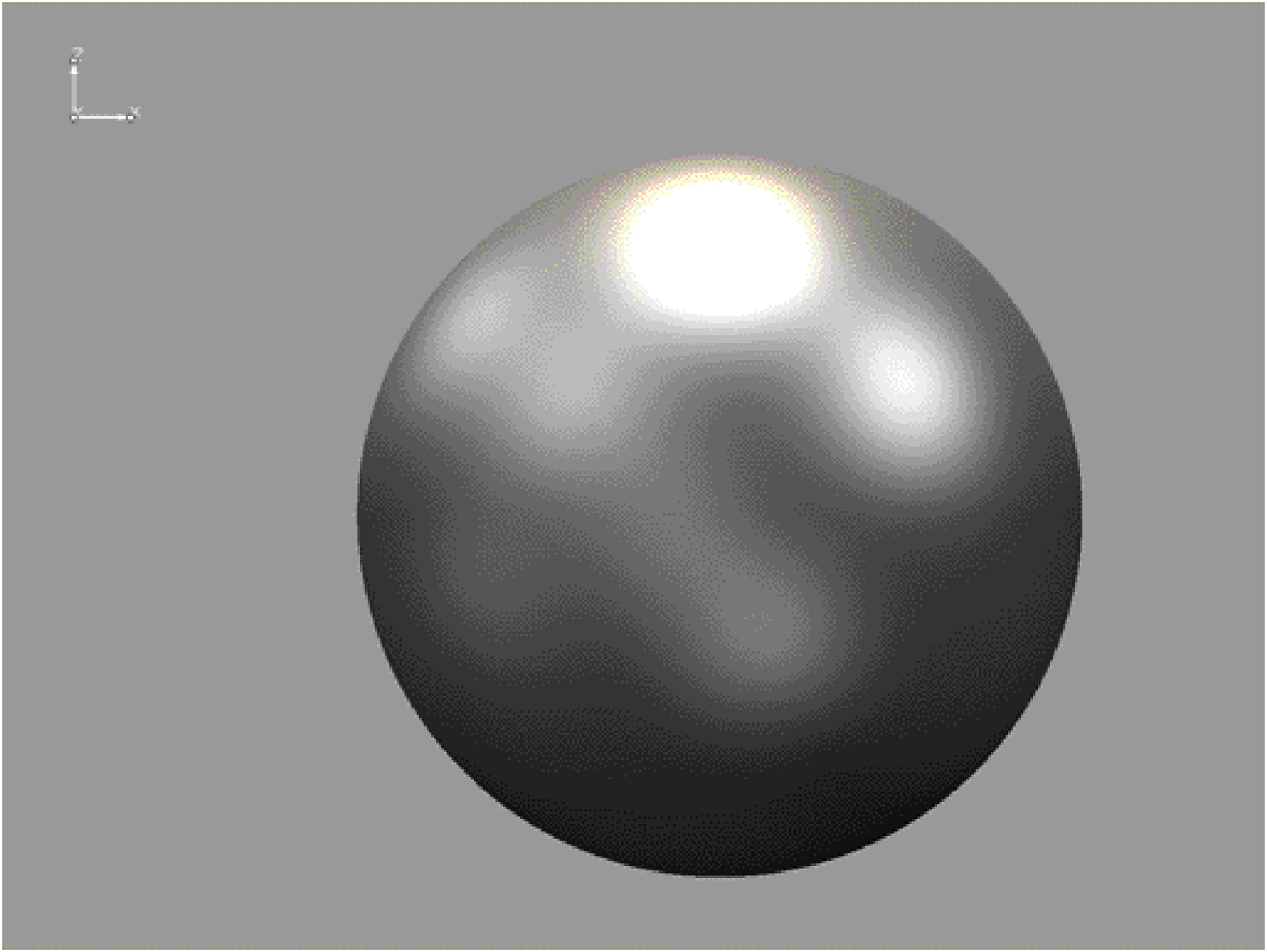}\\
    (e) \includegraphics[width=.6\linewidth]{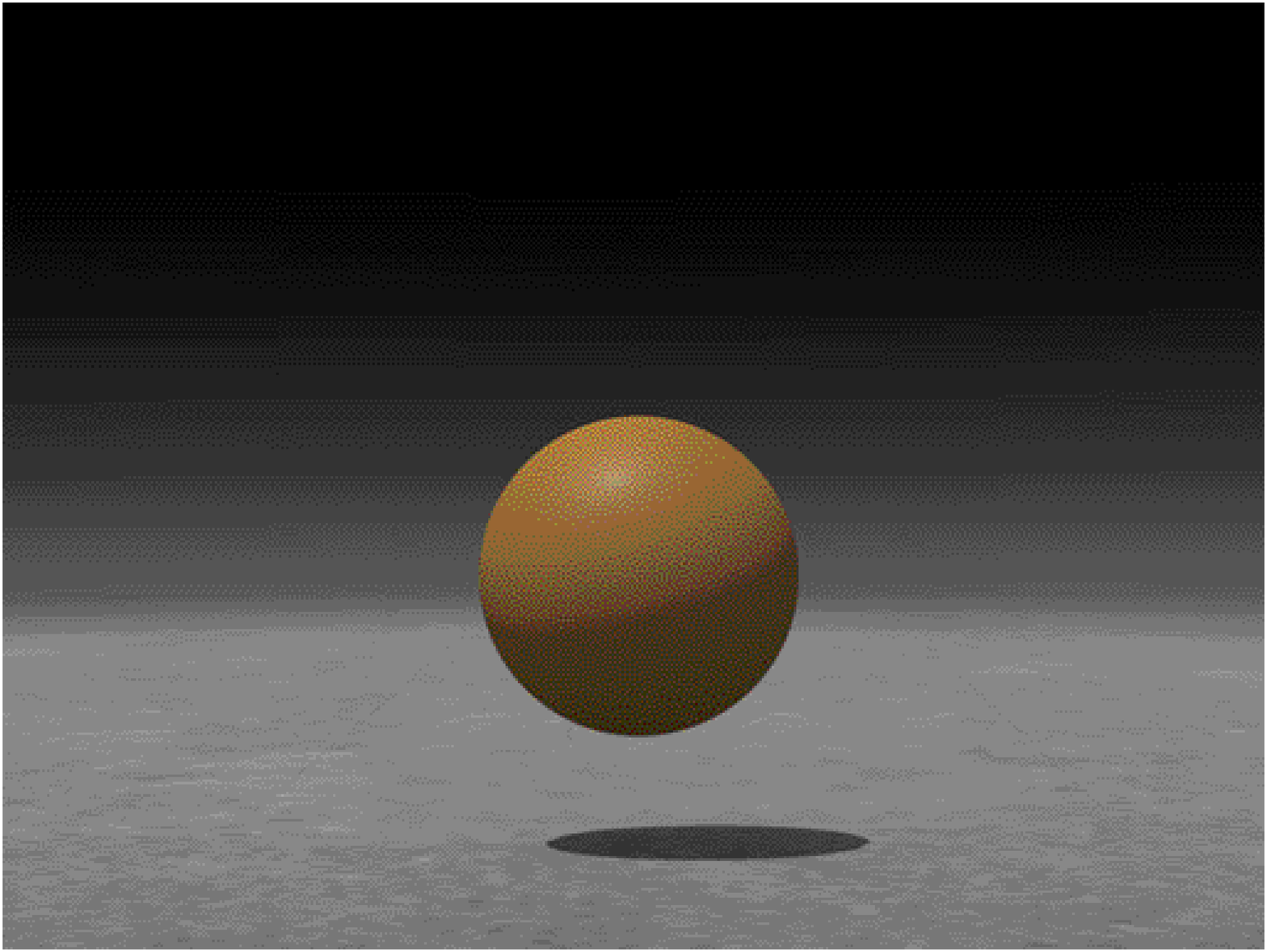} 
    }
  \end{minipage}
  \begin{minipage}[b]{.48\linewidth}
    \centering{
    (a) \includegraphics[width=.6\linewidth]{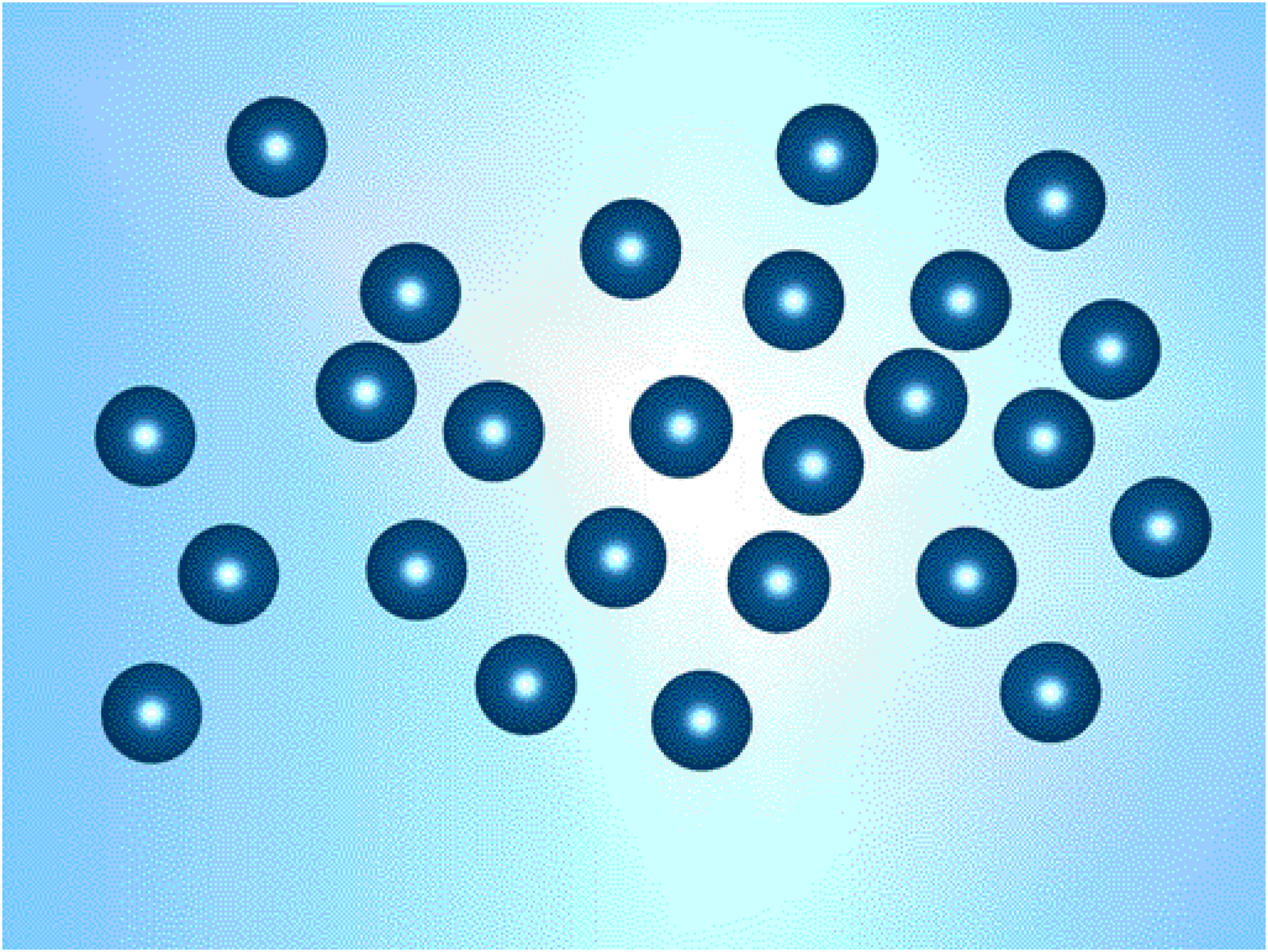} \\
    (b) \includegraphics[width=.6\linewidth]{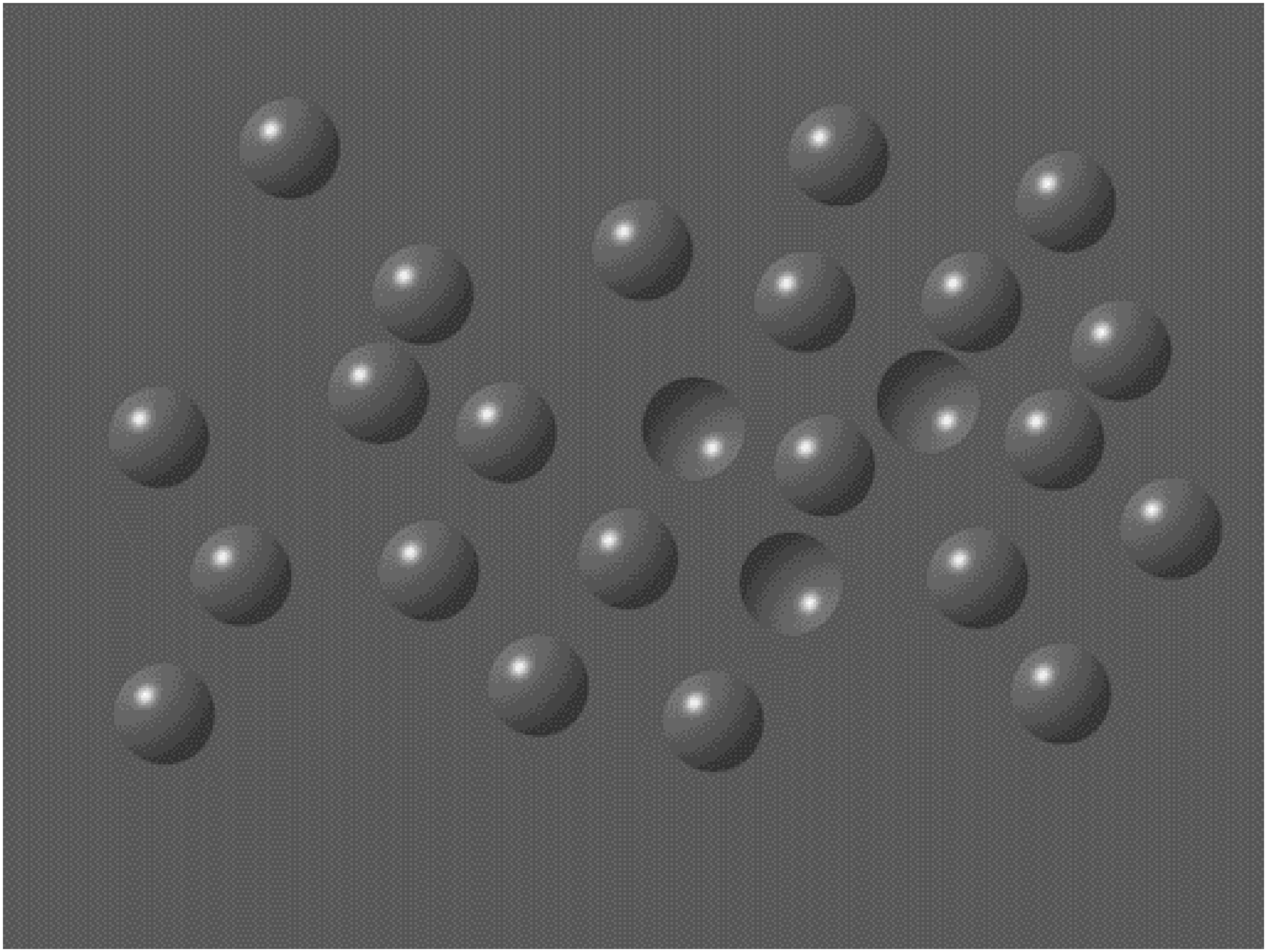} \\
    (c) \includegraphics[width=.6\linewidth]{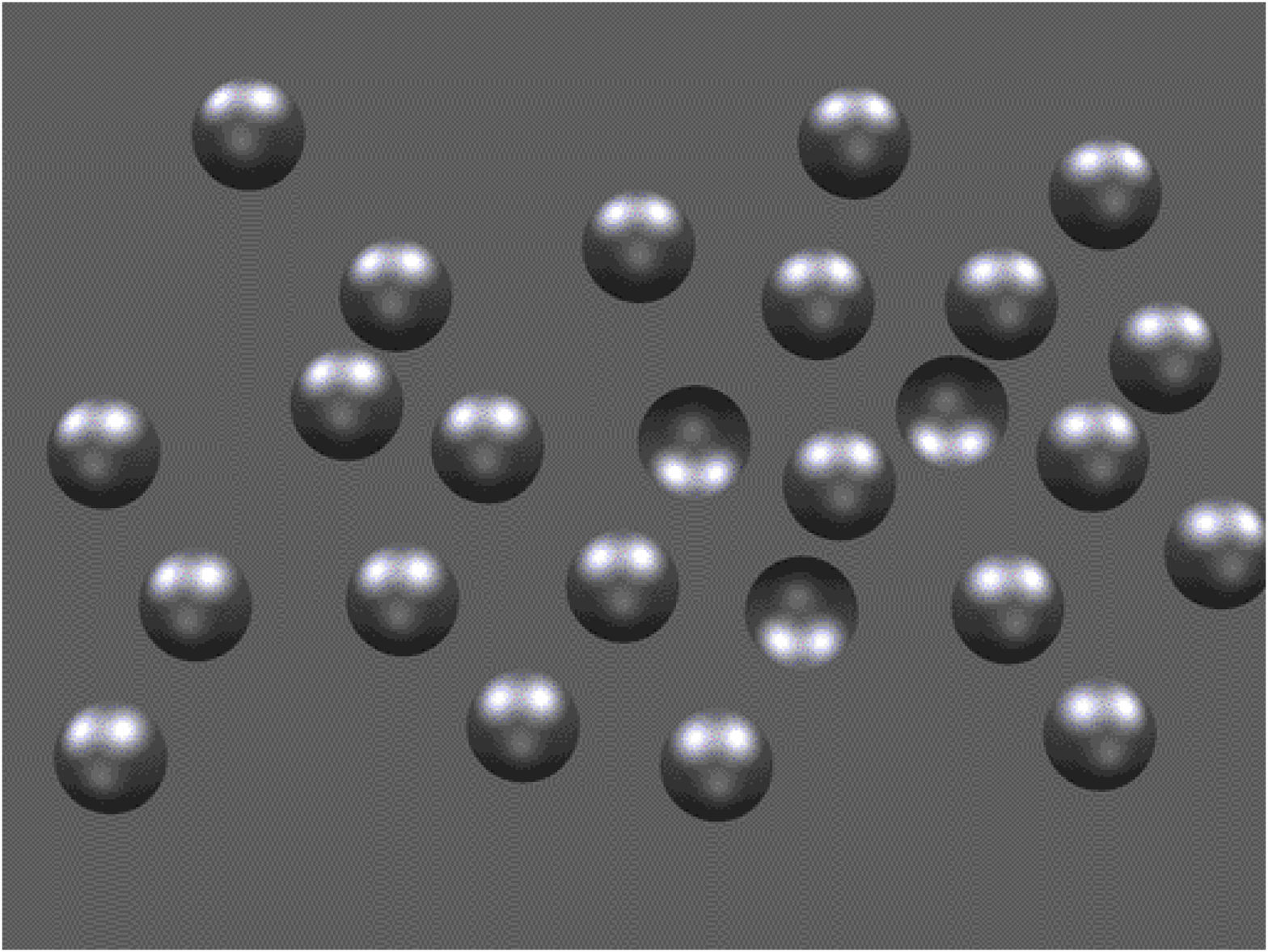} \\
    (d) \includegraphics[width=.6\linewidth]{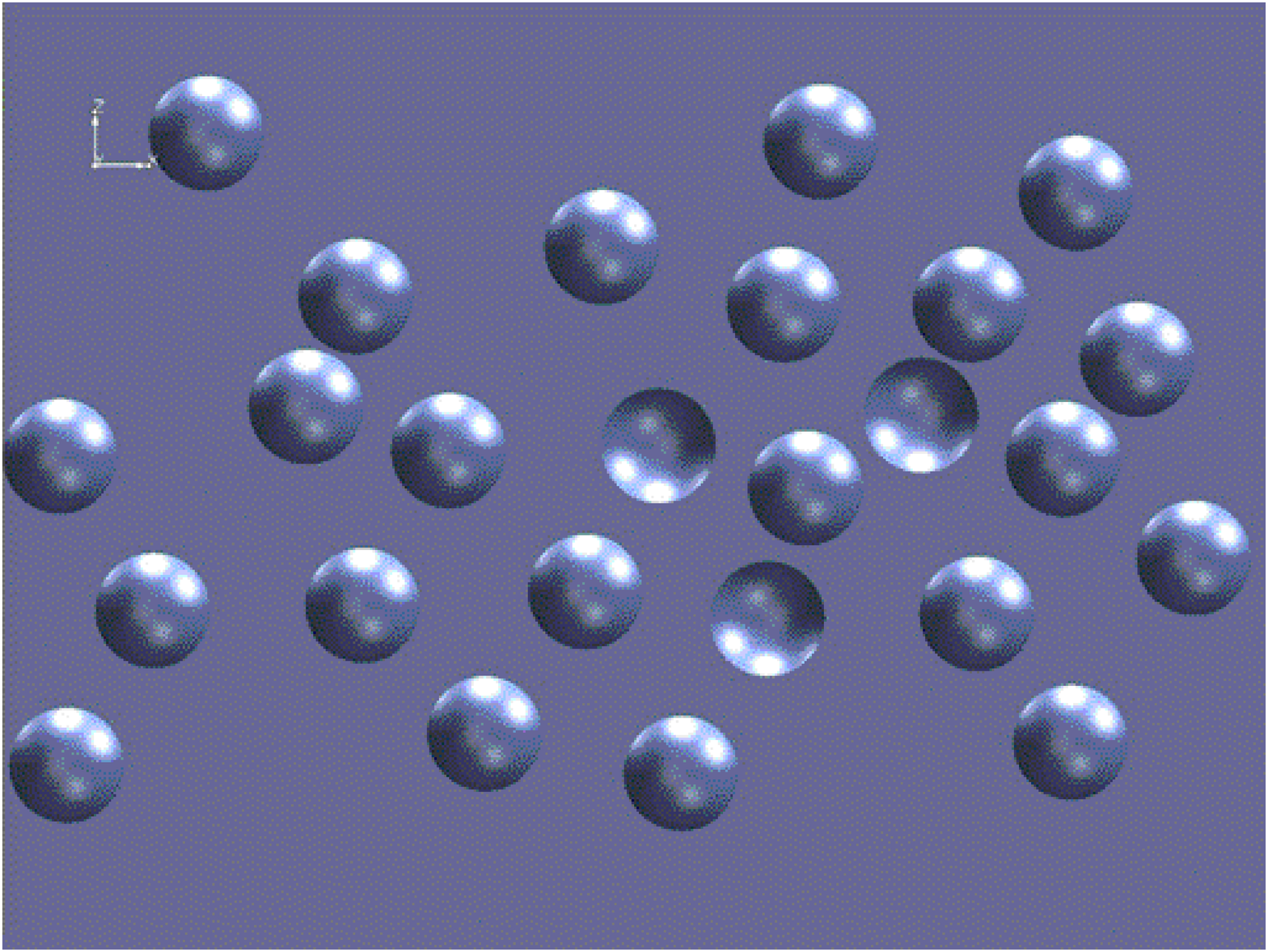} \\
    (e) \includegraphics[width=.6\linewidth]{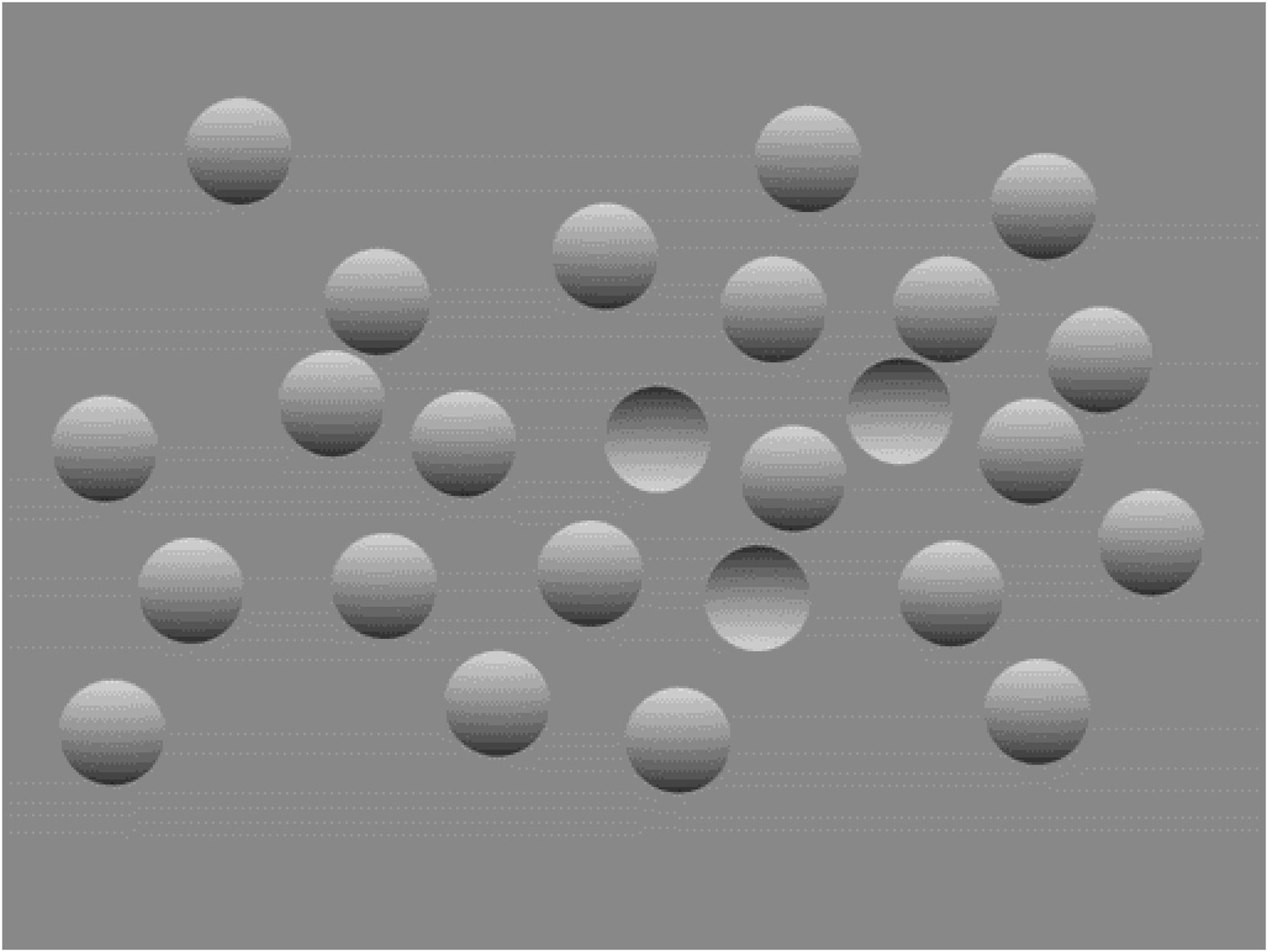}
    }
  \end{minipage}
  \begin{minipage}[t]{.47\linewidth}
    \caption{Balls.
    Renderings of a sphere in some CAD-systems.
    (a) is the default illumination of Alias SurfaceStudio, a single headlight.
    (b) is the default illumination of Alias AutoStudio,
    a single directional light together with an ambient light.
    (c) is a ball under the standard illumination of DBView, 
    a headlight, two additional directional lights and a reflection map.
    (d) is the standard illumination of ICEM Surf, with 8 directional lights.
    (e) shows a rendering of a matte sphere (clay)
    calculated with a ray tracing system that supports a sky illumination.
    }
    \label{fig:balls}
  \end{minipage}
  \begin{minipage}[t]{.47\linewidth}

    \caption{Bumps.
    A Ramachandran pattern rendered under different illumination.
    (a) is the default illumination of Alias SurfaceStudio, a single headlight.
    (b) is the default illumination of  CATIA,
    a single directional light together with an ambient light.
    (c) is the standard illumination of DBView, 
    a headlight, two additional directional lights and a  reflection map.
    (d) is the standard illumination of ICEM Surf, with
    8 directional lights.
    (e) is the illumination by the overcast sky of \cite{MoonSpencer},
    here realized as a graphics shader in DBView/veo.
    }
    \label{fig:bumps} 
 \end{minipage}
\end{figure}

The selection of a single favorite from a set of candidate models
is a typical task in the design process.
Thus the designers were shown digital renderings of a sphere,
see fig.~\ref{fig:balls}.

The task was to select the \emph{best sphere}.
Every designer could stick two bullets on a board for his favorite\footnote{
Such voting is used in business meetings, although it is a scientifically
insufficient, due to possible cross-influences
}.

A sphere was chosen as stimulus
because everybody knows what the shape is meant to look like
and thus can comment 
how his percept of a picture matches this imagined shape.
A second reason was that it
can serve as a probe for the illumination, 
i.e. the light sources, although not seen directly in the picture,
can be explained from their reflection on the sphere.
However, the sphere has certain properties (symmetries, curvature, contour)
that make it unique and therefore 
care must be taken to draw conclusions for generic shapes.

The pictures were rendered in several CAD-systems with their typical or default
settings for the illumination.
The selection represents many of the typical illumination situations
with Phong light sources:
directional lights, headlight and ambient light.
Some systems restrict the type and number of light sources,
e.g. Alias SurfaceStudio has only a fixed headlight,
CATIA has only one or two directional lights,
while Alias AutoStudio, ICEM Surf and DBView/veo can have many light sources,
possibly restricted by the graphics hardware (usually at most eight). 
One picture uses also a reflection map.
In addition, one picture was rendered in a raytracing system with  
global illumination by a sky hemisphere.   

The results are listed in table~\ref{table:ranking}, column balls.

The headlight illumination 
and also the single directional light 
(with an ambient contribution,
that makes the lower left half look flat) 
ranked worst.  

Also the illumination by directional lights with a reflection map,
that is normally used and accepted to show a car,
ranked poorly.
A designer described this illumination \emph{as unreal, like in a thunderstorm}.

A better result was obtained by the illumination with many light sources.
However, one designer described the fig.~\ref{fig:balls}d. 
with the many light sources 
as \emph{a bumpy plate}.
The strong highlights may be (mis-)interpreted as bumps.

The image with the sky illumination was ranked best.
It can be argued that this result is not only accounted for the illumination,
since this picture was the only picture that contained 
a cast shadow of the ball on the floor.

All pictures from CAD-systems show 
overflows in the highlights 
(spots that are too bright),
a typical effect of poor adjustment of light and material parameters.
These spots look flat and their transition is too abrupt
(except with Alias SurfaceStudio), giving an impression of a punched hole.

\begin{table}[htb]
  \captionabove{Scores of different illumination situations,
    given by a group of 10 designers.}
  \label{table:ranking}
  \begin{tabular}[t]{r l c c}
    \hline\noalign{\smallskip}
    Fig. & Illumination                       & balls  & bumps \\
    \hline\noalign{\smallskip}
    a. & headlight                            &  0     &   0   \\ 
    b. & single directional light w/ ambient  &  0     &   4   \\ 
    c. & directional lights w/ reflection map &  1     &   2   \\ 
    d. & many directional lights              &  5     &   1   \\ 
    e. & skylight                             & 13     &   13  \\ 
    \noalign{\smallskip}\hline
  \end{tabular}
\end{table}

\subsection{Detection of concave or convex regions - bump or indentation}

In a second task, 
inspired by \cite{Ramachandran},
the designers were presented images
of a rectangular plate with hemispherical bumps or 
indentations\footnote{
In fact, the bumps and indentations were not complete hemispheres
but rather polar caps with a polar angle of $ 80^0 $. 
This was chosen to avoid a silhouette (occluding contour) to appear in case 
the plate was slightly rotated (which was not done in the task). 
}, 
see fig.~\ref{fig:bumps}.


Again the pictures were rendered with mainly the same CAD-systems and
illuminations as in  fig.~\ref{fig:balls}, 
except that the illumination with directional light and ambient
(fig.~\ref{fig:bumps}b) was rendered with CATIA V5 
and the sky illumination (fig.~\ref{fig:bumps}e)
is the overvcast sky of \cite{MoonSpencer},
realized as a programmable shader in DBView/veo.

The task was to select the image where the indentations could be recognized best.
Again, every designer could stick two bullets on a board for his favorite,
and the results are listed in  table~\ref{table:ranking}, column bumps.

The illumination with three directional lights and reflection map was ranked poorly.

With the headlight, bumps and indentations are indistinguishable.

The illumination with many lights from different directions 
makes it difficult to distinguish bumps and indentations.

The single directional light
(with an ambient contribution) performed better in this task.  

Again, the sky illumination ranked best.
Although it was noted that the bumps and indentations appeared relatively 
flat\footnote{
In view of \cite{JohnstonPassmore} this is not surprising since there
was no specular component.
}.   

This experiment accords with fundamental results of psychophysics,
namely
the prior for light from above \cite{SunPerona},
\cite{KleffnerRamachandran}
and the advantage of diffuse lighting \cite{LangerBuelthoff}.
The CAD-systems ranked poorly since
illumination in CAD-systems is at odds with these priors.

Diffuse illumination from above could be realized by 
an irradiance environment map, cf.  \cite{AkenineMoellerHaines} 
or by implementing an irradiance function, 
e.g. the irradiance of the sky of \cite{MoonSpencer}\footnote{
It seems that in computer graphics only the luminance function 
is recognized 
while the illuminace has been paid little attention. 
}.

\section{Effects of Phong illumination and form appraisal}
This section lists some effects of Phong illumination that are observed by
designers and interfere with form appraisal.

\subsection{Collimated light - night}
The light sources emit collimated light only.
The directional light source and the point light source resp.
illuminates only one half of an object,  
while the other appears entirely black \cite{Birn}.

There are always regions that are not reached by light, and
strong contrasts between dark and bright regions.
This are the lighting conditions of the night,
illuminated by just the moon or some street lamps.

This probably is one reason why so many digital renderings show dark scenes
or even a black background.
And it has caused several strategies for illumination in
CAD-systems that attempt to circumvent this problem, but which can impact form appraisal.

\paragraph{Headlight}
The only directional light source 
that illuminates all visible surface regions in the scene is one in or behind
the camera  -- 
the 'headlight' used in many CAD-systems.
In DBView/veo a headlight is part of the default lights and is switched on by default, 
in Alias SurfaceStudio it is in fact the only light source.
Under a headlight the whole object appears \emph{flat}, 
a fact well known to photographers,
cf. \cite{HunterFuqua},
and that should be avoided,
cf. \cite{Birn}.
However, even in psychophysical experiments a headlight is used sometimes,
e.g. \cite{RodgerBrowse}.

\paragraph{Ambient light}
To avoid black regions, often an ambient contribution is used.
Ambient light makes objects look \emph{flat}, even kinks are flattened out.
Because of its unrealistic effect (cf. e.g. \cite{KoenderinkShading}),
it is hardly used by visualization experts
\cite{Birn}.
However, in \cite{OpenGL}
it comes with a default intensity of $ 20\% $
and CATIA comes even with $ 50\% $. 

The combination of a directional light source (that sends light from above)
with an ambient light gives fairly good results in the 2 tasks of the workshop.

\begin{figure}[ht]
  \begin{minipage}[b]{.49\linewidth}
    \includegraphics[width=.95\linewidth]{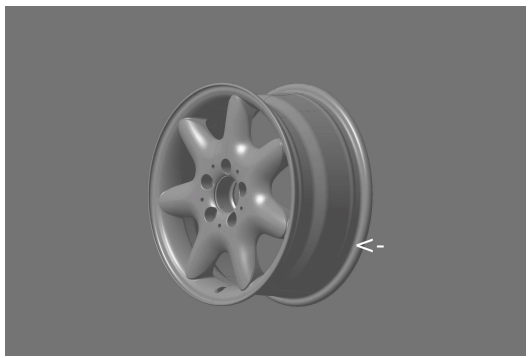}
  \end{minipage}
  \begin{minipage}[b]{.5\linewidth}
    \caption{A rim illuminated by ambient light.} 
    \label{fig:rim}
      Picture rendered in CATIA.
      The hump seems to vanish below the arrow, where only ambient light contributes.
  \end{minipage}
\end{figure}

However, this sort of illumination may lead to serious deception of form,
cf. \cite{Birn} 
or look at fig.~\ref{fig:rim}
for an example from automotive design.

\paragraph{Many lights}
Placing many light sources in the scene is another strategy
to overcome dark regions.
Often these light sources are located on the side of the observer
(front lights) and move with the camera.
This is the default in ICEM Surf.

This illumination makes it difficult to distinguish bumps and indentations.

\paragraph{Moving lights}
\label{sec:movinglights}
In an interactive session, where the model is rotated or the observer walks
around, the headlight, or more general, 
the front lights must move with the camera.

Lights moving with the observer may seem acceptable,
if the observer moves horicontally.
However, with vertical movement or with steeper viewing angle as the observer
approaches, this may look strange to a designer 
as he does not expect 
the brightness of surface regions to change so dramatically 
if seen from above. 
It should also be noted that moving lights are in conflict with the visual prior that
light sources are stationary,
cf. \cite{KerstenMamassianKnill}.

However, the need for moving light may also aid form appraisal.
With only one moving light source the light break moves over the surface
giving strong shape hints.
Such dynamic illumination is used frequently by modellers for surface inspection.

\subsection{Shadows}
The first thing apparent to designers is the absence of cast shadows.
Designers are often puzzled \emph{that light passes through objects} 
without casting a shadow.
Without a shadow, the car model seems to float above the ground,
gap lines (e.g. around the doors) can not be evaluated,
and undercuts (e.g. of the dashboard)
 are underestimated and therefore tend to become exaggerated in
digital work. 

However, the absence of cast shadows is beneficial to headlight.
For other lights, the absence of shadows is visible,
while the latter is correct even without shadows.

\subsection{Reflections}
Reflection maps also show something in areas that are not reached by light.
Therefore they often play an undue role in computer graphics.
Designers do not like too strong reflections, 
they say that the model sometimes \emph{is mirrored to death} 
or ask to \emph{turn off the reflections, I want to see the form}.

However, reflections can also give valuable shape hints, cf. \cite{Fleming}.

\subsection{Highlights}
Highlights play an important role in shape recognition, 
and many designers prefer fair glossy materials like metallic car paint
to appraise form.  They say
\emph{metallic car paints emphasize the form}.

Glossy materials show stronger gradients of brightness 
along curvature 
than matte materials.
The glossy highlights emphasize the ridges and fillets on a surface.
Along a ridge the variation of the normal is greater than in regions with
smaller principal curvatures
(cf. e.g. \cite{KoenderinkShape})
With the normals also the reflected viewing directions 
subtend a larger solid angle. 
Now, given a distribution of light sources, 
it is more likely that the highlight from a light source will happen to lie on
the ridge.
And if the highlight is on the ridge, it will be distorted,
i.e. elongated in the direction of the ridge, 
cf. \cite{BlakeBuelthoff}, \cite{BlakeBrelstaff}.
Thus ridges become visible, and designers 
can read them as \emph{feature lines or character lines}.

\paragraph{Size of highlight} 
\label{sec:highlightsize}
When asked to comment on the appearance of highlights in digital renderings,
a designer called them \emph{too soft and dull}.

\begin{figure}[ht]
  \begin{minipage}[b]{.49\linewidth}
    \includegraphics[width=\linewidth]{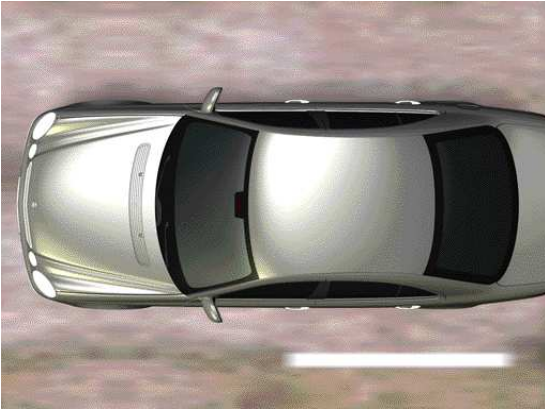}
  \end{minipage}
  \begin{minipage}[b]{.5\linewidth}
    \caption{Sharpest Phong highlight on a car roof.}
    \label{fig:highlight} 
    Picture rendered in DBView.
  \end{minipage}
\end{figure}

In OpenGL, the shininess exponent of the Phong highlight is restricted to 7 Bit
(i.e. $ m_{shiny} \le 127 $).  
This causes large highlights.
Fig.~\ref{fig:highlight} shows the sharpest Phong highlight on a car roof.
In reality, the highlight of the sun on a smooth convex surface can always be covered
by the tip of the small finger at arm length.

The larger a highlight is on a ridge, the softer the shape appears.
Crisp highlights are needed to convey crisp shapes\footnote{
It is interesting to note that designers use the same words 
- \emph{soft} and \emph{hard} -
to describe material properties (roughness)
and shape properties (curvature).
}.
To achieve a more realistic sun highlight, the shininess exponent would have to
go up to 5-10 000.

On the other hand, the larger highlights compensate for the 
discrete character (zero width) of the Phong light sources, 
see sec.~\ref{sec:highlightsize},
or may somewhat simulate the glare effect of bright highlights in reality.

\paragraph{Highlight cutoff}

\begin{figure}[htb] 
  \begin{minipage}[b]{.49\linewidth} 
    \includegraphics[width=\linewidth]{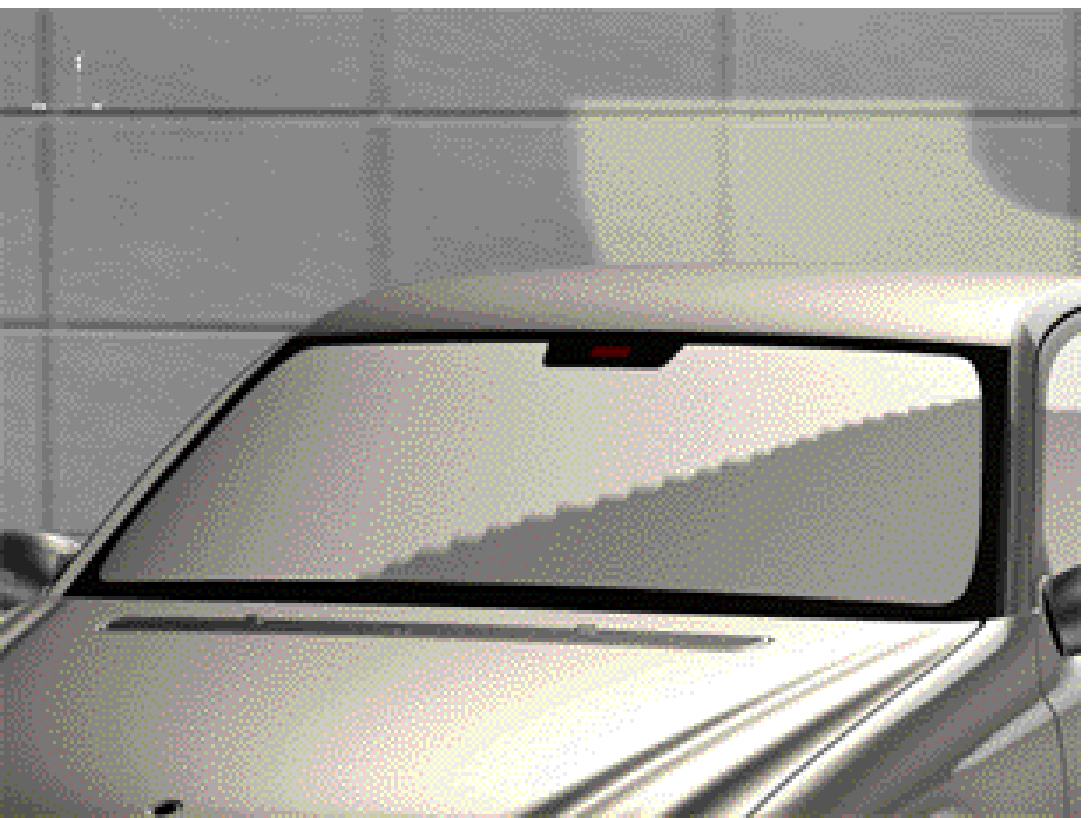}
  \end{minipage}
  \begin{minipage}[b]{.5\linewidth}
  \caption{Highlight cutoff on a windscreen.}
    Picture rendered in ICEM Surf.
  \label{fig:highlightcutoff}
    \end{minipage}
\end{figure}
  
An artifact noted by designers are the \emph{jagging light breaks} 
sometimes seen around Phong specular highlights,
see fig.~\ref{fig:highlightcutoff}.
This is the highlight cutoff described by 
\cite{WooPearceOuellette}.
What may seem as a tesselation problem is in fact the discontinuity
of the Phong specular reflection function
along the terminator, 
the silhouette line with respect to the light direction, 
see \cite{AkenineMoellerHaines}.  
In contrast to the diffuse reflection component, 
which continuously goes down to zero at grazing angles following Lamberts
cosine law,
the specular component is independent of the angle between surface normal 
and light direction as long as this angle is acute,
and suddenly vanishes as the angle gets obtuse.

Highlight cutoff  may be seen if a light source 
is nearly opposite to the viewing direction.
It does not occur with headlights or (moving) frontlights.
Thus it is concealed by poor lighting. 

The modified Phong model of \cite{Lewis} is a simple cure to this flaw.

Highlight cutoff is a nuisance that may appear suddenly in an interactive
form appraisal session.
It is clearly identified as an artefact if jagging is concerned
and may question the credibility of digital form appraisal.  
It may be even more critical, if the tesselation is so fine that the cutoff
line is smooth.  
In that case the cutoff line may be interpreted as a surface feature,
possibly as a kink line,
but more likely, the region behind the cutoff is seen as a shadow.

Although it is physically unrealistic and 
doesn't behave like a shadow if viewed in motion (it rather is an isophote),
a moving terminator may even leverage form appraisal,
similar to sec.~\ref{sec:movinglights}.

\section{Behaviour of Light and Material}

Designers state that \emph{adjusting lights and materials is too complicated}.
Illumination is an  \emph{enduring task},
in particular in an interactive 3D  application where the user may
walk  around the object and see it from behind.
The lighting often needs re-adjustment for a new perspective.

Firstly this needs an expertise that designers do not need in the real world
- designers hardly ever adjust lights when looking at real objects\footnote{
The notable exception is class A surface inspection,
where panels with parallel light rods are positioned around the model 
to follow the reflection lines.
}.
Secondly, it is so complicated,
because \emph{the lights behave differently to the real world}.

\subsection{Illumination terminator } 
A drawback of illumination in CAD-systems is that
it is normally not gamma-corrected.
I.e. the non-linear transfer function from coding to brightness on the display
is not taken into account, cf. \cite{PoyntonGamma}, \cite{PoyntonHDTV}.
And this effect is visible to designers.

\begin{figure}[htb]
  \includegraphics[width=0.49\linewidth]{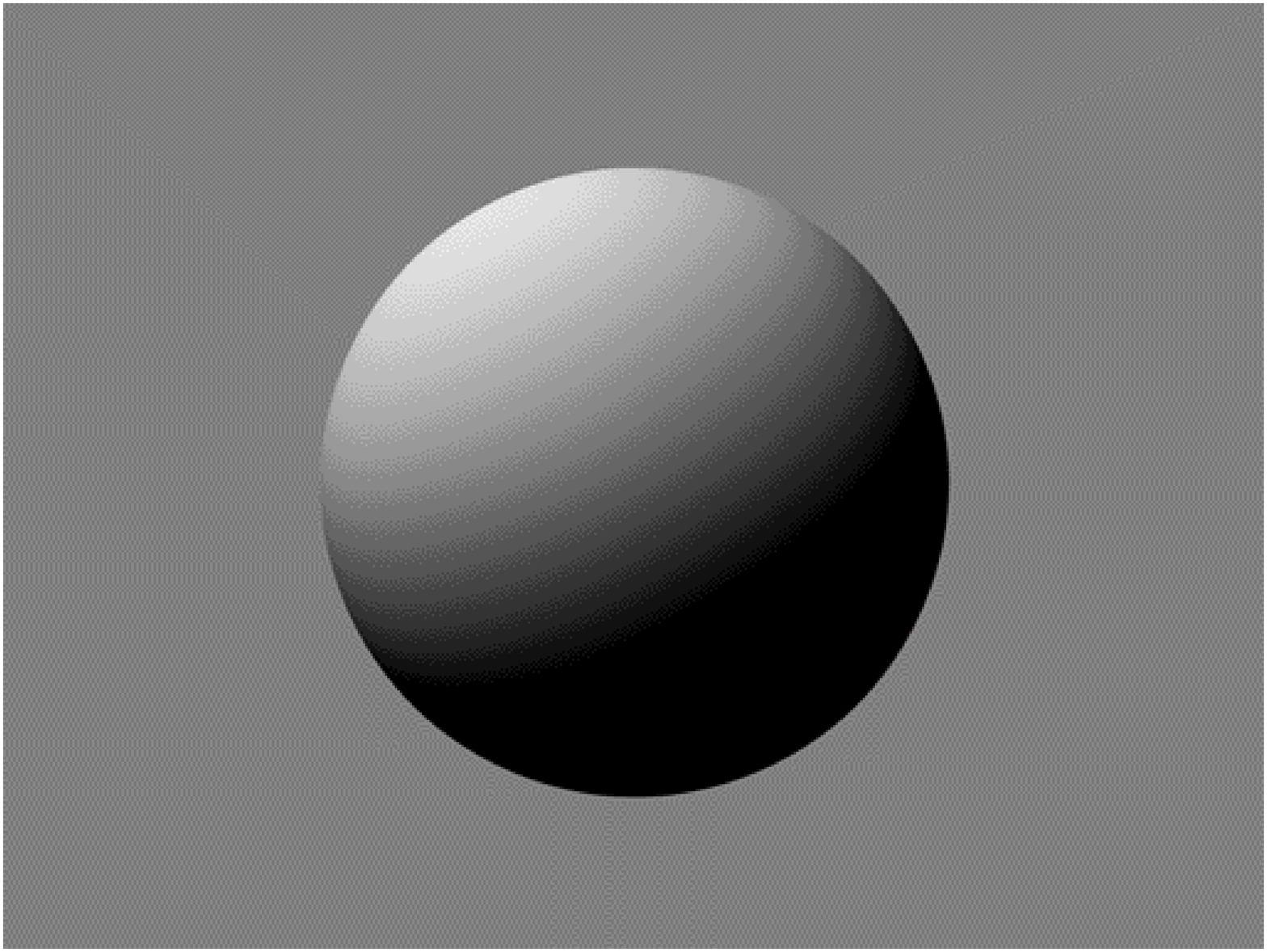}
  \includegraphics[width=0.49\linewidth]{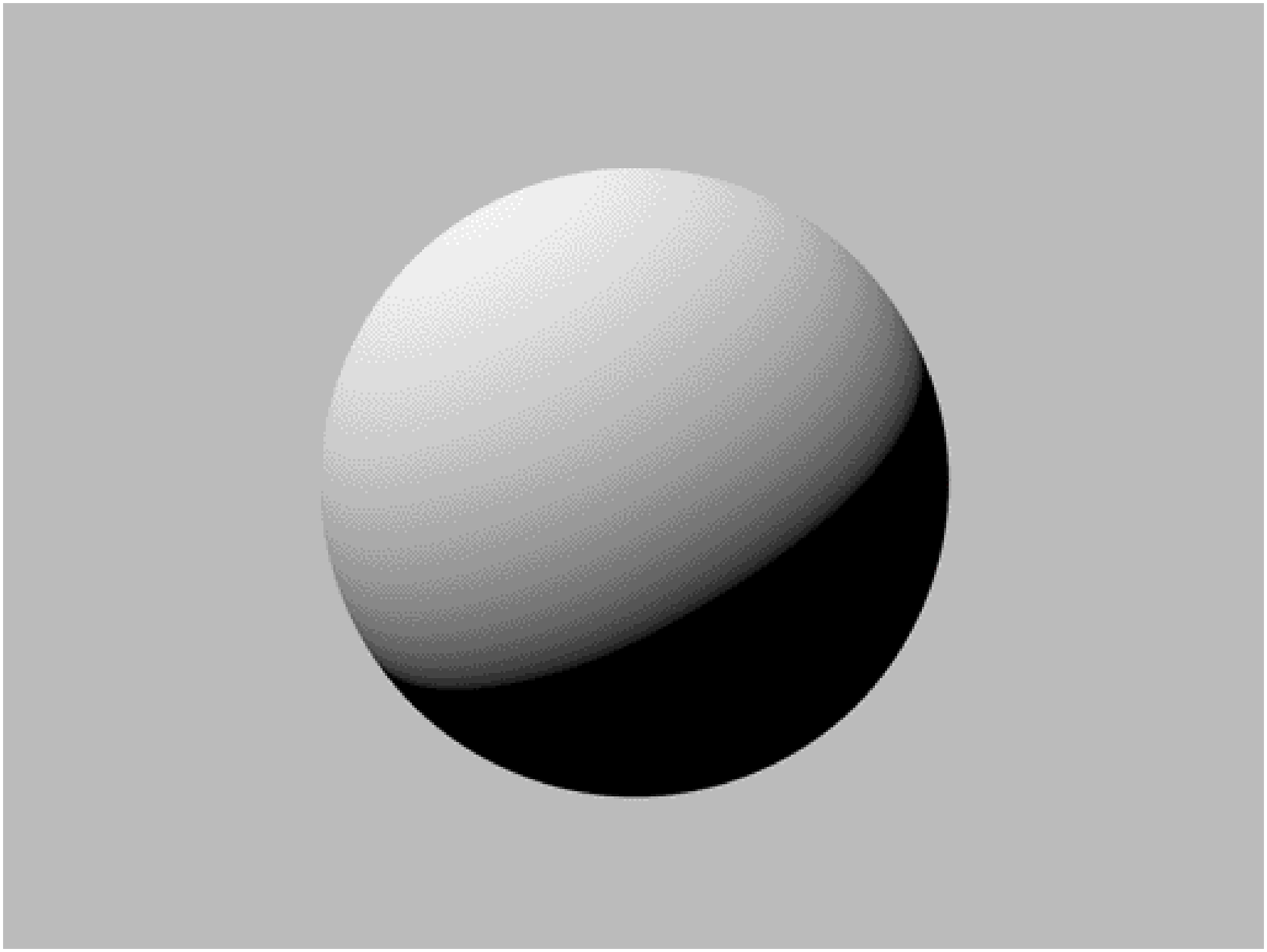}
  \caption{A sphere illuminated by a directional light source.}
  \label{fig:gamma}    
  Left without and right with gamma correction.
  The sphere was assigned an ideal matte white material (Lambertian shader).
  Pictures rendered with Alias AutoStudio. 
\end{figure}

Fig.~\ref{fig:gamma} shows two pictures of a sphere illuminated by a directional
light source, calculated
without and with gamma correction, resp.

The two pictures were presented to a group of designers.
When asked, which picture showed a ball illuminated by a directional light
source, most designers named the first picture.

However, when asked, which picture showed a ball illuminated by a "laser beam",
the majority voted for the second picture.

Physically, it is clear that a directional light source 
represents an ideal laser beam,
and should give rise to a hard light break a.k.a. terminator, cf. \cite{Birn}.

Surprisingly, CAD-systems often show soft light breaks.
This may look realistic, since real life light sources
also produce soft light breaks.
But the position of the terminator is wrong:
The area illuminated by a real light source would extend over $ 90^0 $
(polar angle)
while with the ideal light source it is strictly confined to a hemisphere.

Correcting the gamma transfer
makes light breaks get harder and look more unrealistic,
i.e. reveals the laser beams.
This could give a reason, why gamma correction 
- although well known -
is missing in CAD-systems.

\subsection{Superposition of light}
Another manifestation of gamma is the superposition of light.
From the mathematical point of view, it is just simple:
summation and power function do not commute: 
$ (1 + 1)^2  \not= 1^2 + 1^2  $.   

\begin{figure}[ht]
  \includegraphics[width=\linewidth]{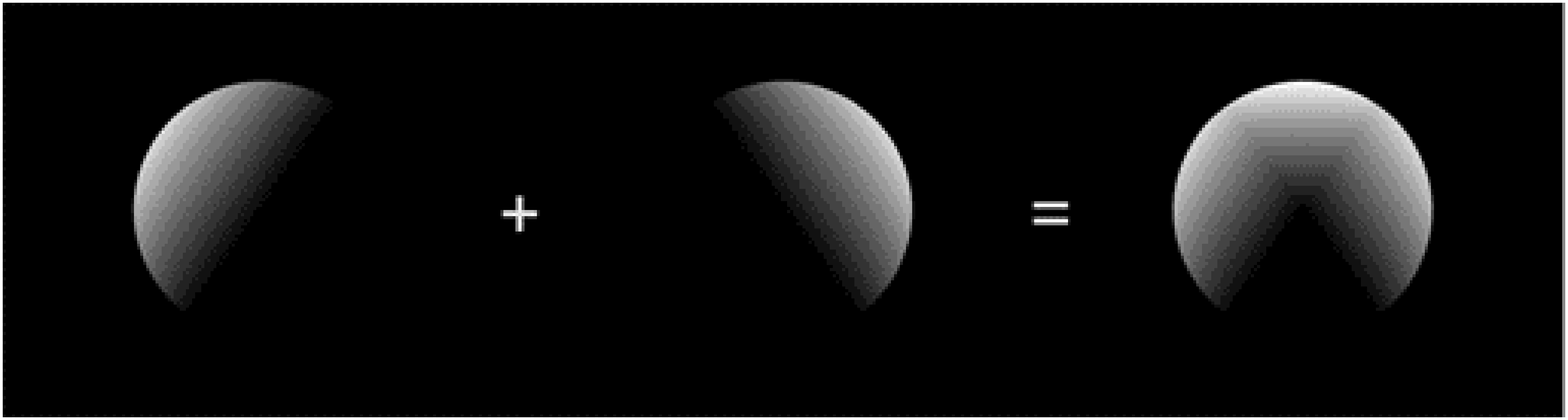}\\[1mm]
  \includegraphics[width=\linewidth]{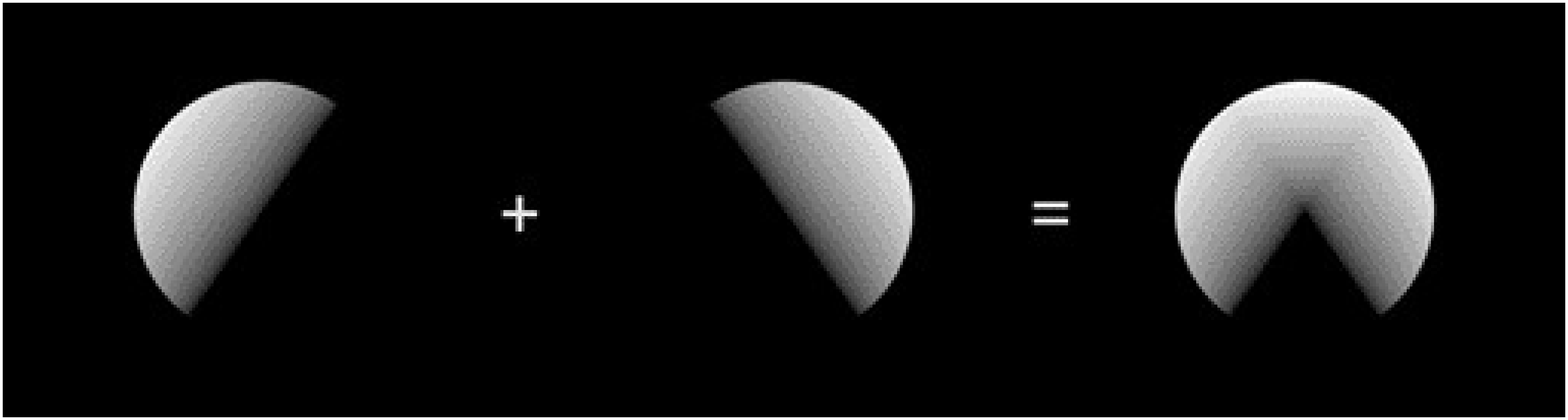}
  \caption{Superposition of light.}
  \label{fig:Superposition} 
  Each picture shows 3 balls, 
  of an ideal matte white material (Lambertian shader),
  illuminated by directional light sources.
  The ball on the right is illuminated by two directional 
  light sources.
  The balls on the left and middle are illuminated by 
  only one directional light source, from the left or the right resp. 
  (a) shows the effect of the lights 
  as seen in many CAD-systems (here rendered with Alias AutoStudio),
  The fig. also exhibits strong terminators
  as upward pointing diagonal lines in the right picture.
  These might be (mis-)interpreted as a kink on the surface,
  (b) shows the effect with a physically correct summation 
  as rendered in Alias AutoStudio with a gamma correction 
  suitable for display gamma of $ 2.2 $
  (AutoStudio allows a gamma correction in its batch rendering).
\end{figure}

Fig.~\ref{fig:Superposition} shows the superposition of light as a symbolic formula,
one calculated without and one with gamma correction.

Again, the two pictures fig.~\ref{fig:Superposition}
were presented to a group of designers.
The pictures were presented on a display with standard sRGB settings
\cite{sRGB},
i.e. display gamma $ = 2.2 $ .
Designers found it \emph{strange and unrealistic} 
that two dim lights should sum up to such a
bright light in the middle.
The gamma corected superposition looked plausible to them.

In CAD-systems running on ordinary screens\footnote{
The screens are not linearly calibrated. 
Also
linear calibration would be a bad thing, 
since this looses the benefits of perceptually uniform encoding,
see \cite{PoyntonGamma}.
}, 
the superposition of light is too bright.
This is one reason 
that overflows occur so frequently.
Loosely speaking,
with a display gamma of roughly 2, 
CAD-systems compute 
"$  1 + 1 = 4 $".
Probably these are the only systems engineers use nowadays 
that calculate that way.

The strange behaviour of superposition, 
the unexpected appearance of bright spots or overflows,
is one reason that makes adjusting light so complicated.

Proper gamma correction would reconcile the situation.

\subsection{Coupling of light and material}
Another effect that annoys designers who want to adjust light or material
parameters, is the coupling of light and material.

\paragraph{Material gloss and intensity of light source}
One well known effect is that shininess is not normalized, 
cf. e.g. \cite{Lewis}.
Modifying the shininess exponent 
to adjust the sharpness (concentration) of the highlight 
(i.e. the gloss of the material)
also varies the amount of reflected light or the intensity of the light source.
If the shininess exponent gets smaller, more light is reflected.
For shininess exponent $ m_{shiny}< 6 $ more light is reflected than received!

This causes frequent overflows for small shininess exponent 
and the overflow region appears flat.

\paragraph{Material gloss and size of light source}
Another coupling effect is due to the discrete nature of Phong light sources.
The shininess exponent of Phong is not only used to adjust the gloss of a
material but also to adjust the apparent size (spatial dimension) of a
positional light source, cf. \cite{Birn}.

\begin{figure}[htb]
  \begin{minipage}[b]{.49\linewidth}
    \includegraphics[width=.95\linewidth]{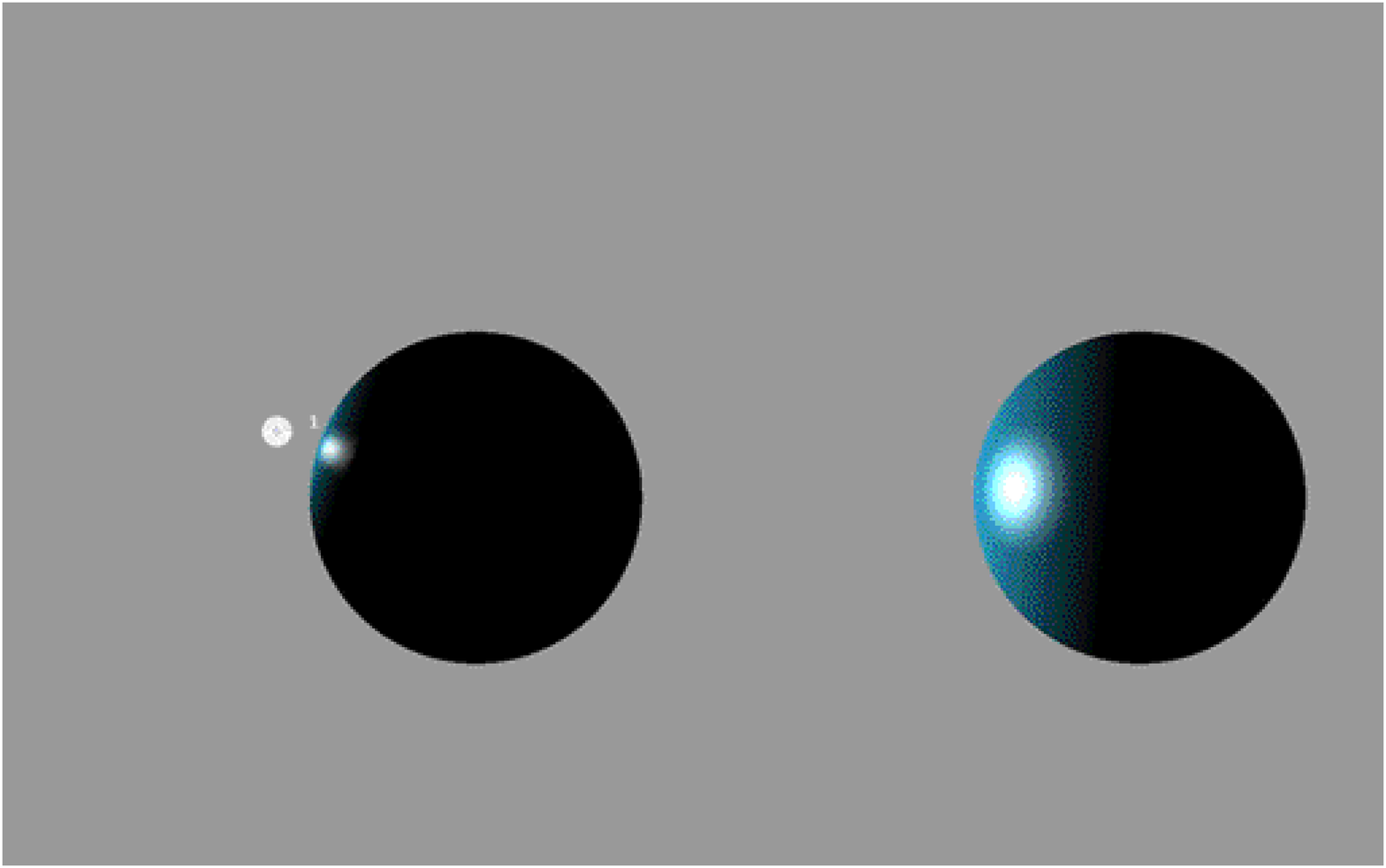}
  \end{minipage}
  \begin{minipage}[b]{.5\linewidth}
     \caption{Two balls illuminated by a point light source.}
     \label{fig:2balls} 
     Picture rendered in ICEM Surf.
  \end{minipage}
\end{figure}

Fig.~\ref{fig:2balls} shows two balls illuminated by a point light source.
The shininess exponent of the balls was carefully adjusted 
such that the size of the highlight on the ball close to the light source
matches the apparent size of the light source as indicated by the light symbol.
The highlight on the right ball is bigger than on the left,
while in reality, an extended light source would cause 
a highlight on the distant ball that is smaller than on the nearby 
ball\footnote{
Taking into account the distance attenuation 
(which is usually ignored in CAD-systems) 
it should also be weaker.
}.
Within the Phong illumination model, this appearance could only be achieved,
if the distant ball was assigned a different shininess exponent,
i.e. another material, or with per object lighting.
  
Thus a material parameter (gloss or size of highlight)
accounts for properties of the light source (intensity and size).

\begin{table}[hbt] 
  \captionabove{Effects of Phong illumination, impact and palliatives.
    The possible impacts on appearance or form appraisal are listed
    together with effects or practices that conceal or alleviate these impacts.}
  \label{table:effects} 
  \begin{tabular*}{\linewidth}{p{.2\linewidth}  p{.3\linewidth}  p{.45\linewidth}}
    \hline\noalign{\smallskip}
    effect  & impact (appearance)   &  palliative effects and counter strategies
    \\
    \noalign{\smallskip}\hline
    collimated light only     & night illumination   
    &  headlight, ambient, many lights,  \newline
    moving lights, reflection map, context \newline
    (dark environment, black background)
    \smallskip\\
    headlight                & flat 
    &   no shadows, \newline  context (dark background)
    \smallskip\\
    ambient light            & flat     & texture, reflection map  
    \smallskip\\
    many lights              & confuses (concave-convex \newline discrimination)
    &   complexity of scene
    \smallskip\\
    moving lights            & unnatural behaviour   &
    \smallskip\\
    \noalign{\smallskip}\hline
    no cast shadows          & no depth     &    night, headlight
    \smallskip\\
    \noalign{\smallskip}\hline
    highlight size           & edges softer     &  collimated light  
    \\                       & glare     & low dynamic range display 
    \smallskip\\
    highlight cutoff         & (jagging) light break
    &  headlight, moving front lights    
    \smallskip\\
    highlight overflow       & flat
    \smallskip\\
    \noalign{\smallskip}\hline
    illumination terminator  & soft light breaks
    & gamma transfer of display 
    \smallskip\\
    superadditivity \newline of superposition            
    & overflows (flat), \newline  unnatural behaviour     
    &  complexity  
    \smallskip\\
    coupling light-material  & unnatural behaviour, \newline complicated usage  
    &  per object lighting
    \smallskip\\
    \noalign{\smallskip}\hline
  \end{tabular*}
\end{table}

\section{Summary and Conclusion} 

The Phong illumination model has one main deficiency and several faults,
that are listed in table~\ref{table:effects}.
There is no diffuse light.
It is the diffuse light of sky that makes the difference between day and night,
not that the night was just darker. 
Thus, the illumination in CAD-systems
is \emph{in essence the illumination of the night}.

Some effects make objects appear flat,
others exhibit light breaks that are not physically plausible
and might be taken for a kink.
Together these may give the impression of a
\emph{cardboard box},
as a designer had put it.
Thus Phong illumination is not just a limited simulation of reality,
it has a bias in a particular direction.

Other effects 
are in conflict with  real life experience or
are at odds with priors of visual perception.
And this presumably has an impact on form appraisal.

Some effects are 
concealed by others
or are alleviated by common practices of digital rendering.
Probably these palliatives have prevented faults from being fixed.

It seems that Phong illumination is an unintentionally balanced model
with well known faults,
but where luckily one flaw conceals another one.
Thus curing one flaw only, may make it worse.

Nevertheless it is a question,
why such a poor illumination is still alive in CAD-systems.
Is it
because perception can well adapt to poor lighting\footnote{ 
or as a modeller has put it:
\emph{the screen lies but you can learn}.
},
or is it that shading is a weak cue,  
in particular if the user can interact with the digital model and see it in motion,
or is it just its simplicity?

It is a challenge for research
to define an illumination model
that aids visual perception
and is still simple enough to enter CAD-systems.

\bibliographystyle{apalike}
\bibliography{Literatur}

\end{document}